\documentclass[11pt]{article}

\input epsf
\usepackage{amsmath}

\usepackage{amssymb}
\usepackage{theorem}

\usepackage{enumerate}

\usepackage{mathrsfs}

\usepackage{geometry}
\usepackage{graphicx}
\usepackage{epstopdf}

\numberwithin{equation}{section}

\theoremstyle{plain}

\newtheorem{proposition}{Proposition}

\theoremstyle{definition}

\theoremstyle{remark}
\newtheorem{remark}{Remark}

\def\R{\ensuremath{\mathbb R}}
\def\N{\ensuremath{\mathbb N}}

\def\I{\ensuremath{{\bf 1}}}
\def\e{{\ensuremath{\rm e}}}

\def\p{\ensuremath{\mathbb P}}

\def\o{\ensuremath{\underline{\omega}}}

\def\1{{\bf 1}}

\def\eps{\varepsilon}

\newcommand{\qand}{\quad\text{and}\quad}

\def\p{\ensuremath{\mathbb{P}}}

\numberwithin{equation}{section}

\begin{document}
\bibliographystyle{plain}

\title {SAMPLING LOCAL PROPERTIES OF  ATTRACTORS VIA
 EXTREME VALUE THEORY}

\author{Davide Faranda \thanks{Laboratoire SPHYNX, Service de Physique de l'Etat Condens\'e, DSM, CEA Saclay, CNRS URA 2464, 91191 Gif-sur-Yvette, France. e-mail: $<$davide.faranda@cea.fr$>$.}
\and Jorge Milhazes Freitas \thanks {Centro de Matem\'{a}tica \& Faculdade de Ci\^encias da Universidade do Porto, Rua do
Campo Alegre 687, 4169-007 Porto, Portugal. e-mail:$<$jmfreita@fc.up.pt$>$.}
\and Pierre Guiraud \thanks{CIMFAV, Facultad de Ingenier\'ia, Universidad de Valpara\'iso, Valpara\'iso, Chile. e-mail: $<$pierre.guiraud@uv.cl$>$.}
\and Sandro Vaienti
\thanks{Aix Marseille Universit\'e, CNRS, CPT, UMR 7332, 13288 Marseille, France and
Universit\'e de Toulon, CNRS, CPT, UMR 7332, 83957 La Garde, France.
e-mail:$<$vaienti@cpt.univ-mrs.fr$>$.}}
\maketitle

\begin{abstract}
We provide formulas to compute the  coefficients entering the affine scaling needed to get a non-degenerate function for the asymptotic distribution of the maxima of some kind of observable computed along the orbit of a randomly perturbed dynamical system. This will give information on the local geometrical properties of the stationary measure. We will consider systems perturbed with additive noise and with observational noise. Moreover we will apply our techniques to chaotic systems and to contractive systems, showing that both share the same qualitative behavior when perturbed.
\end{abstract}

\section{Introduction}
A general problem in dynamical systems theory is to give a quantitative characterization of the limiting invariant sets like  attractors or repellers, whose properties are essential to understand the behavior of complex systems. In the last years, the results of the Extreme Value Theory (EVT) have brought new techniques that allow to quantify the geometrical and dynamical properties of a certain class of systems. In the case of  absolutely continuous invariant measures (acim), precise analytical results can be obtained in terms of classical Extreme Value Laws (EVLs) and depend on the fulfillment of general mixing conditions and on the observables considered. In fact, those observable are designed in such a way that \textit{extreme events} are equivalent to detect the recurrence of an  orbit in a neighborhood  of a given point in the phase space. A collection of such events, under appropriate renormalization, is distributed according to one of the three classical EVLs, namely the Gumbel, the Frechet and the Weibull distributions. The values of the normalizing constants are linked to the local behavior of the measure  and, provided  the dynamics is chaotic and the measure is absolutely continuous, they depend only on  the number of extremes extracted and on the phase space dimension. Several difficulties  arise whenever  singular invariant measures are considered. In \cite{LFTV, FV2} this problem was addressed almost numerically and  a few analytic results have been exhibited  in \cite{FFT, HNT, HVR, GHN}.

Let us now explain in detail where the just indicated problem is  and how we could deal with it by introducing random perturbations: this will constitute the first main contribution of this paper. At this regard we need to come back to basics and introduce the theory. Let us therefore suppose that $(Y_n)_{n\in\N}$ is a sequence of real-valued random variables defined on the probability space $(\Psi, \mathbb{P}).$ We will be interested in the distribution of the maximum $M_n:=\max\{Y_0, Y_1, \dots, Y_{n-1}\}$ when $n\rightarrow \infty.$ It is well known that the limiting distribution is degenerate unless one proceed to a suitable re-scaling of the levels of exceedances. The precise formulation is the following:
 we have an Extreme Value Law  for $(M_n)_{n\in\N}$ if there is a non-degenerate distribution function
  $H:\R\to[0,1]$ with $H(0)=0$ and,  for every $\tau>0$, there exists a sequence of levels $(u_n(\tau))_{n\in\N}$ such that
\begin{equation}
\label{eq:un}
\lim_{n\to\infty}n\mathbb{P}(Y_0>u_n)\to \tau,
\end{equation}
and for which the following holds:
\[
\lim_{n\to\infty}\mathbb{P}(M_n \le u_n)\rightarrow 1-H(\tau).
\]

The motivation for using a normalizing sequence $(u_n)_{n\in\N}$ satisfying \eqref{eq:un} comes from the case when $(Y_n)_{n\in\N}$ are independent and identically distributed. In this i.i.d.\ setting, it is clear that $\mathbb{P}(M_n\leq u)= (F(u))^n$, being $F(u)$ the cumulative distribution function for the variable $u$. Hence, condition \eqref{eq:un} implies that
\[
\mathbb{P}(M_n\leq u_n)= (1-\mathbb{P}(Y_0>u_n))^n\sim\left(1-\frac\tau n\right)^n\to\e^{-\tau},
\]
as $n\to\infty$.  Note that in this case $H(\tau)=1-\e^{-\tau}$ is the standard exponential distribution function. Let us now choose the sequence $u_n=u_n(y)$ as the one parameter family $u_n=y/a_n+b_n$, where $y\in \mathbb{R}$ and $a_n>0,$ for all $n\in \mathbb{N}$. Whenever the variables $Y_i$ are i.i.d. and  for some constants $a_n>0$, $b_n\in\R$, we have $\mathbb{P}(a_n(M_n-b_n)\leq y)\rightarrow G(y)$, where the convergence occurs at continuity points of $G$, and $G$ is non-degenerate, then $G_n$ will converge to one of the three EVLs: Gumbel, Fr\'echet or Weibull.  The law obtained depends on the the common distribution of the random variables, $F$.

When $Y_0,Y_1,Y_2,\ldots$ are not independent, the standard exponential law still applies under some conditions on the dependence structure. These conditions will be stated in detail later and  they are usually designated by $D_2$ and $D'$; when they
 hold for  $(Y_n)_{n\in\N}$ then there exists an extreme value law for $M_n$ and $H(\tau)=1-e^{-\tau},$ see Theorem 1 in \cite{FF}. We want to stress that these two conditions alone do not imply the existence of an extreme value law;  they require, even to be checked, that the limit (\ref{eq:un}) holds. It turns out that for the kind of observables we are going to introduce, and which are related to the local properties of the invariant measure, the limit (\ref{eq:un}) is difficult to prove when the invariant measure is not absolutely continuous, since one needs the exact asymptotic behavior of that measure on small balls. Instead it turns out that whenever the system is randomly perturbed, the limit (\ref{eq:un}) is more accessible and in particular it will be given by a closed formula in terms of the strength of the noise, see Proposition 1 and 2 below.  Moreover that formula could be used in a reversed way (in the following we call this procedure {\em inverting the technique}) : since the sequence $(u_n)_{n\in\N}$ is now uniquely determined for any $n$, a numerical sampling for $(u_n)_{n\in\N}$ which provides convergence to the extreme value law, will bring information on the local geometrical  properties of the stationary measure: this approach was successfully used, for instance, in \cite{FV1, FV2}. \\

We already showed in a preceding article \cite{FFTV} that random perturbations of regular systems, in particular rotations, induce the appearance of extreme value laws since the perturbed systems acquires a chaotic behavior. We pursue, and this is the second main issue of this paper,  the same objective here by considering two different kinds of stochastic perturbations of (piecewise) contracting maps. The first will be given by additive noise and in this case our analysis will be mostly numerical. The second one will be a sort of (rare) random contamination of a deterministic orbit, and in this case we will announce and state  complete analytic results for the determination of the limit (\ref{eq:un}) first, and for the successive checking of the conditions $D_2$ and $D'$.

\section{Random dynamical systems}
In this section we will introduce the two ways of perturbing a given dynamical system, the random transformations and the observational noise.
\subsection{Random transformations}

Let us consider  a sequence of i.i.d. random variables $(W_k)_{k\in \N}$ with values $(\omega_k)_{k\in \N}$ in a space $\Omega_{\varepsilon}$ and with common probability distribution $\theta_{\varepsilon}$. Let $X\subset\R^d$ be a compact set equipped with the Lebesgue measure $m$ defined on the Borel $\sigma$-algebra, and $(f_\omega)_{\omega\in \Omega_{\varepsilon}}$ a family of measurable transformations such that $f_{\omega}:X\to X$ for all $\omega\in \Omega_{\varepsilon}$\footnote{In the following when we will refer to a dynamical system $(X,f,\mu)$ we will mean that $f$ is defined on $X$ and preserves the Borel probability measure $\mu;$ if we will  write $(X,f)$, this will simply  correspond  to the action of $f$ on $X.$ }. Given a point $x\in X$ and a realization $\underline{\omega}= (\omega_1,\omega_2,\dots)\in \Omega_{\varepsilon}^\N$ of the stochastic process $(W_k)_{k\in \N}$, we define the random orbit of $x$ as the sequence $(f^n_{\underline{\omega}}(x))_{n\in\N}$, where
\[
f^0_{\underline{\omega}}(x)=x\quad\text{and}\quad f^n_{\underline{\omega}}(x)=f_{\omega_n}\circ f_{\omega_{n-1}}\circ\cdots\circ f_{\omega_1}(x)\qquad\forall n\geq 1.
\]
The transformations $f_{\omega}$ will be considered as stochastic perturbations of a deterministic map $f$ , in the sense that they will be taken in a suitable neighborhood of $f$ whose {\em size} will be determined by the value of $\eps$, see below.  We could therefore  define a Markov process on $X$ with transition function
\begin{equation}\label{gre}
 L_{\varepsilon} (x, A)=\int_{\Omega_{\varepsilon}}\mathbf{1}_{A}(f_{\omega}(x))d\theta_{\varepsilon}(\omega),
 \end{equation}
where $A\in X$ is a measurable set,  $x\in X$ and $\mathbf{1}_{A}$ is the indicator function of the set $A$. A probability measure $\mu_{\varepsilon}$ is called  \textit{stationary} if for any measurable set  $A$ we have:
$$
\mu_{\varepsilon}(A)=\int_{X} L_{\varepsilon}(x,A)d\mu_{\varepsilon}(x)
.$$
We call it an absolutely continuous stationary measure ({\em acsm}), if it has a density with respect to the Lebesgue measure. \\

Given a map $f:X\to X$, we will consider two kind of random perturbations. The first one is the additive noise, which corresponds to the family $(f_\omega)_{\omega\in \Omega_{\varepsilon}}$ of random transformations defined by
\[
f_{\omega}(x)=f(x)+\omega\quad\forall\,x\in X.
\]
In this case each $\omega$ belong to the hypercube  $\Omega_{\varepsilon}\subset\R^d$ of side $2\eps$  centered at zero, and equipped with the measure $\theta_{\varepsilon}=\frac{m}{(2\varepsilon)^d}\I_{\Omega_\eps}$, which is the normalized Lebesgue measure restricted to $\Omega_{\varepsilon}$.  For these perturbations, some additional assumptions may be necessary to ensure that the image of each $f_\omega$ is included in $X$.

Notice that for additive noise,
\begin{equation}\label{KK}
 L_{\varepsilon} (x, A)= \theta_{\varepsilon}(\omega\in \Omega_{\varepsilon}: f(x)+\omega\in A)\le \frac{m(A)}{(2\varepsilon)^d}
\end{equation}
which implies that if the stationary measure exists for such random transformations, it is absolutely continuous w.r.t. the Lebesgue measure on the ambient space. In the paper \cite{benedicks2006random} Benedicks and Viana proved that for additive noise the stationary measure exists and is unique for the H\'enon map. The same remains true for the Lozi map and these two maps will be studied in Sections \ref{SECLOZI} and \ref{SECHENON}.\\

The second kind of random transformations we will consider have been introduced by Lasota and Mackey (see \cite{LM}, for instance) and correspond to \em randomly applied stochastic perturbations\em. They  consist in operating an aleatory reset of the initial condition of the dynamical system $(X,f)$ at each failure  of a Bernoulli random variable: if $(x_{n})_{n\in\N}$ denotes the successive states of such a random dynamical systems, then at each time $n\in\N$ we have $x_{n+1}=f(x_n)$ with probability $(1-\epsilon)$ and $x_{n+1}=\xi_n$ with probability $\epsilon$, where $\xi_n$ is the realization of a random variable with value in $X$. This kind of perturbation corresponds to the family $(f_\omega)_{\omega\in \Omega_{\varepsilon}}$ of random transformations defined by
\begin{equation}\label{RASP}
f_{\omega}(x)=\eta f(x) + (1-\eta)\xi\qquad\forall\, x\in X,
\end{equation}
where $\omega=(\eta,\xi)$ is a random vector with value in $\Omega_{\varepsilon}=\{0,1\}\times X$. The  two components $\eta$ and $\xi$ of $\omega$ are independent and $\eta$ is a Bernoulli variable with the probability of being $0$ equal to $\eps$, while $\xi$ is a random variable that we will suppose Lebesgue-uniformly distributed on $X$. The joint distribution $\theta_\varepsilon$ of these two components is the product of the Bernoulli measure with weights $(1-\eps, \eps)$ and the uniform measure on $X$.

We notice that in this case the transition function $L_{\eps}(x,\cdot)$ will give a singular measure for any $x$, since for any measurable set $A$
\[
L_{\eps}(x,A)= \theta_{\varepsilon}(f_{\omega}(x)\in A)= \eps m(A)+ (1-\eps)\delta_{f(x)}(A)
\]

\subsection{Observational noise}

A different type of perturbation is the  {\em observational noise}. Here, the noise affects the observations $(y_n)_{n\in\N}$ of the orbits of a dynamical systems $(X,f)$, but does not affect the dynamics itself. Precisely, it consists in replacing the orbit $(f^n(x))_{n\in\N}$ of a point $x\in X$, by the sequence $(y_n)_{n\in\N}$ defined by
\[
y_n=f^n(x)+\eps \xi_n\quad\forall\,n\in\N,
\]
where $\eps>0$ and  $(\xi_n)_{n\in\N}$ is a sequence of  i.i.d random vectors, which take values in the hypercube of $\mathbb{R}^d$ centered at $0$ and of side $2$, $\Omega_1:=\{u\in \mathbb{R}^d; \ |u|_i\le 1\, i=1,\dots,i\}$,   and with common distribution  $\theta$, which we choose absolutely continuous with density $\rho\in L^{\infty}_{m}$, namely $d\theta(\xi) =\rho(\xi) dm(\xi)$, with $\int_{\Omega_1} \rho(\xi)  dm(\xi)=1$\footnote{Each $\xi$ is a vector with $d$ components; all these components are independent and distributed with common density $\rho'$; the  product of such marginals $\rho'$'s gives  $\rho$.}. \\

\section{Level sets for the EVL }\label{LEVEL}
In the following we consider a dynamical system  $(X,f,\mu)$ perturbed   with the noises introduced in the previous section. In order to study the extreme value statistics, we define a stochastic process $(Y_n)_{n\in\N}$ by composing a given observable $\phi:X\to\R$ respectively with:
\begin{itemize}
 \item random transformations, that is $Y_n=\phi\circ f^n_{\underline{\omega}}$ for all $n\in\N$. In this case  $(Y_n)_{n\in\N}$ will be a stationary process if we consider the probability $\mathbb{P}=\mu_{\varepsilon}\times \theta_{\varepsilon}^{\mathbb{N}}$, which corresponds to the {\em annealed} situation where we average over the initial condition and over the realization of the noise.

     \item  observational noise, that is $Y_n=\phi\circ(f^n +\varepsilon \xi_n)$ for all $n\in\N$. In this  case $(Y_n)_{n\in\N}$ will be a stationary process if  $\mathbb{P}=\mu\times \theta^{\mathbb{N}},$ where $\mu$ is the invariant measure for $f$. This measure is defined on the product space $X\times \Omega_1^{\mathbb{N}}$ with the product $\sigma$-algebra. A point in this space will be the couple $(x,\ \overline{\xi}:=\{\xi_0,\xi_1,\cdots,\xi_n,\cdots \})\in X\times \Omega_1^{\mathbb{N}}$.
\end{itemize}
In both cases we will chose the observable $\phi(x)=-\log||x-z||$, where $||\cdot||$ is the euclidean  norm on $X$ and $z$ a point of $X$. As we anticipated in the Introduction, such an observable is related to recurrence in small sets, since the distribution of the maximum of the random observable $\phi \circ f^k_{\o}, k=0,\dots, n-1$  up to the level $u_n$, coincides with the distribution of the first entrance of the random orbit into the ball $B(z,e^{-u_n})$ centered at $z$ and of radius $e^{-u_n}$

We now state the two conditions which ensure weak dependence of the process and which allow to get the limiting distribution of the maxima.\\

\noindent\textbf{Condition}[$D_2(u_n)$]\label{cond:D2} We say that $D_2(u_n)$ holds for the sequence $Y_0,Y_1,\ldots$ if for all $\ell,t$
and $n$,
\[
|\mathbb{P}\left(Y_0>u_n\cap  \max\{Y_{t},\ldots,Y_{t+\ell-1}\leq u_n\}\right)-\mathbb{P}(Y_0>u_n)
  \p(M_{\ell}\leq u_n)|\leq \gamma(n,t),
\]
where $\gamma(n,t)$ is decreasing in $t$ for each $n$, and
$n\gamma(n,t_n)\to0$ when $n\rightarrow\infty$ for some sequence
$t_n=o(n)$.

Now, let $(k_n)_{n\in\N}$ be a sequence of integers such that
\begin{equation}
\label{eq:kn-sequence-1}
k_n\to\infty\quad \mbox{and}\quad  k_n t_n = o(n).
\end{equation}

\noindent\textbf{Condition}[$D'(u_n)$]\label{cond:D'} We say that $D'(u_n)$
holds for the sequence $Y_0, Y_1, Y_2, \ldots$ if there exists a sequence $(k_n)_{n\in\N}$ satisfying \eqref{eq:kn-sequence-1} and such that
\begin{equation}
\label{eq:D'un}
\lim_{n\rightarrow\infty}\,n\sum_{j=1}^{\lfloor n/k_n \rfloor}\mathbb{P}( Y_0>u_n,Y_j>u_n)=0.
\end{equation}
As we said in the Introduction when these two  conditions
 hold for  $(Y_n)_{n\in\N}$ then there exists an extreme value law for $M_n$ and $H(\tau)=1-e^{-\tau},$ but provided  that the limit (\ref{eq:un}) is true.\\
 
 When $Y_0, Y_1, Y_2,\ldots$ are not independent, exceedances of high thresholds may have a tendency to appear in clusters, which creates the appearance of a parameter $\vartheta$ in the exponential law, called the Extremal Index. We say that $Y_0,Y_1,\ldots$ has an \emph{Extremal Index} (EI) $0\leq\vartheta\leq1$ if we have an EVL for $M_n$ with $\bar H(\tau)=\e^{-\theta\tau}$ for all $\tau>0$. We can say that, the EI is that measures the strength of clustering of exceedances in the sense that, most of the times,  
it can be interpreted as the inverse of the average size of the clusters of exceedances. In particular, if $\theta=1$ then the exceedances appear scattered along the time line without creating clusters. The conditions $D_2$ and $D'$ stated above are useful to check the existence of an extreme value law corresponding to an EI equal to $1$. In \cite{FFT2} the authors established a connection between the existence of an EI less than 1 and periodic behaviour and  they gave conditions like $D_2$ and $D'$ which they called $D_p$ and $D'_p$.  Since we are not going to investigate in detail the EI here, we simply observe that the role of balls will be now taken by annuli, in the sense that  the limit law corresponding to no entrances up to time $n$ into the ball $\{Y_0\le u_n\}$ is equal to the limit law corresponding to no entrances into the annulus $\{Y_0>u_n, \;Y_p\leq u_n\}$ up to time $n$, where $p$ is the period of the periodic point $z$.

The next results address the question of finding the scaling sequence $(u_n)_{n\in\N}$ in the case where it is affine $u_n=\frac{y}{a_n}+b_n;$ we will see that under suitable assumptions on the noise, the limit (\ref{eq:un}) is actually an equality at each level $n$ and it provides $\tau=e^{-y}.$ We begin with the following result proved in \cite{FFTV}
\begin{proposition}\label{P1}
 Let us consider the dynamical systems $(X, \mathcal{B}, \mu, f)$, where $X$ is a compact set in $\mathbb{R}^d$, and $\mu$ is a probability invariant measure on the sigma-algebra $\mathcal{B}$. We perturb it with observational noise and we consider the associated  process $X_n(x,\overline{\xi}) :=-\log(||f^nx+\varepsilon \xi_n)-z||)$ endowed with the probability $\mathbb{P}=\mu\times \theta^{\mathbb{N}}.$ We suppose moreover that $\theta$ is the Lebesgue measure measure on $S$. Then the linear sequence $u_n:=y/a_n + b_n$ defined in (\ref{eq:un}) verifies\footnote{We notice that since we are now considering the probability measure on an hypercube, the formulae in \cite{FV1} should be modified  according those presented here by adding the  factors $2^d$ and $K_d$ in the   logarithm.}

\[
 a_n=d \qand b_n=\frac{1}{d}\log\left(\frac{n\ K_d\ \mu(B(z,\varepsilon))}{(2\varepsilon)^d}\right).
\]
\end{proposition}
where $K_d$ is the volume on the unit hypersphere in $\mathbb{R}^d$.
We presented in \cite{FV1} class of systems for which the conditions $D_2$ and $D'$, ensuring the existence of the Gumbel law, could be explicitly  checked. In these cases, the  result given by the preceding proposition shows that the sequence $(b_n)_{n\in\N}$ is determined by the measure of the ball centered at the point $z$ and with positive radius $\eps$. Singular measures behaves often as
 $\mu(B(z, \eps))\approx \eps^D$ where $D$ is  an estimation of the fractal dimension of $\mu$ at the point $z$. Of course $D$ is not the dimension itself, which is obtained in the limit of vanishing $\eps$, but, inverting the technique, its value could be obtained by sampling the $b_n$  at  a fixed intensity for the noise and in order to get the Gumbel's law. The numerical results presented in the next sections  show that the values of $D$ provided by this method are in good agreement with the true Hausdorff dimension of the measure.\\

An analogous  result can be obtained for random transformation given by additive noise $f_{\omega}(x)=f(x)+\omega$, where $\omega$ is now chosen in the  hypercube  $\Omega_{\varepsilon}\subset\R^d$ of side $2\eps$  centered at zero, and equipped with the measure $\theta_{\varepsilon}=\frac{m}{(2\varepsilon)^d}\I_{\Omega_\eps}$.
 We remind  that in this case $\mu_\eps$ is the stationary measure and $\mathbb{P}=\mu_\eps\times \theta^\N$ is the product measure.
\begin{proposition}\label{P2}
 Let us consider  the dynamical systems $(X, \mathcal{B}, f)$, where $X$ is a compact set in $\mathbb{R}^d.$ We  perturb it with random transformations admitting the stationary measure $\mu_{\eps}.$ We consider the associated process $X_n(x,\o) :=-\log(||f_{\o}^n(x)-z||)$ endowed with the probability $\mathbb{P}=\mu_\eps\times \theta_\eps^{\mathbb{N}}.$  Then a suitable normalising linear sequence $u_n:=y/a_n + b_n$ is given by:
 $$
 a_n=d; \ b_n=\frac1d\ \log\left(\frac{ n K_d\,\mu_\eps(f^{-1}(B(\zeta,\eps))}{(2\varepsilon)^d}\right),$$
 \end{proposition}

\noindent{\em Proof:} 

Mimicking the proof of the previous proposition  and using stationarity we set 
 $$u_n:=\frac1d\ \log\left(\frac{ n K_d \,\mu_\eps(f^{-1}(B(\zeta,\eps))}{(2\varepsilon)^d\tau}\right)$$

 Using stationarity of $\mu_\eps$ we can write:

 \begin{align*}
 n\p(X_0>u_n)&= m \p(X_1>u_n)=n\p(\{(x,\o): |f(x)+\omega-\zeta|< \e^{-u_m} \})\\
& =n\iint \I_{B(\zeta-f(x),\e^{-u_m})}(\omega)\I_{\Omega_\eps}d\theta^\N(\o)d\mu_\eps(x)\\&
= \int n\theta_\eps(B(\zeta-f(x),\e^{-u_m})\cap \Omega_\eps)d\mu_\eps(x)
  \end{align*}

Now, by definition of $u_n$ and the fact that $\theta_{\varepsilon}=\frac{m}{(2\varepsilon)^d}\I_{\Omega_\eps}$ we have:
\begin{align*}
n\theta_\eps\Big(B(\zeta-f(x),\e^{-u_m})\cap \Omega_\eps\Big)&\leq  n\frac1{(2\eps)^d} m(B(\zeta-f(x),\e^{-u_m}))=\frac{nK_d\,\e^{-d\,u_n}}{(2\eps)^{d}}\\
&=\frac{\tau}{\mu_\eps(f^{-1}(B(\zeta,\eps))}\leq C_{\eps}\nonumber,
\end{align*}
where $C_\eps>0$ depends on $\eps$ but not on $m$. Then, 
can apply Lebesgue's dominated convergence theorem to the sequence functions $h_n(x)=n\theta_\eps\Big(B(\zeta-f(x),\e^{-u_n})\cap \Omega_\eps\Big)$. Note that since Lebesgue measure is translation invariant,  we can write 
$$
\lim_{n\to\infty}h_n(x)=\lim_{n\to\infty}\frac{nK_d\,\e^{-d\,u_n}}{(2\eps)^{d}}\I_{B(\zeta,\eps)}(f(x))=\frac{\tau}{\mu_\eps(f^{-1}(B(\zeta,\eps))}\I_{f^{-1}(B(\zeta,\eps))}(x),
$$
because $\theta_\eps(\Omega_\eps)=1$ and if $|\zeta -f(x)|>\eps$ the intersection $B(\zeta-f(x),\e^{-u_n})\cap \Omega_\eps$ is eventually empty (which explains the indicator on the rhs).

It follows then by Lebesgue's dominated convergence theorem that:
\begin{align*}
\lim_{n\to\infty} n\p(X_0>u_n)&= \int \lim_{n\to\infty}n\theta_\eps^\N(B(\zeta-f(x),\e^{-u_n})\cap \Omega_\eps)d\mu_\eps(x)\\
&=\int \frac{\tau}{\mu_\eps(f^{-1}(B(\zeta,\eps))}\I_{f^{-1}(B(\zeta,\eps))}(x)d\mu_\eps(x)=\tau.
\end{align*}
Hence, $\lim_{n\to\infty}\p(M_m\leq u_m)=\e^{-\tau}$. Moreover, since $u_n:=y/a_n + b_n$, taking $y=-\log(\tau)$, and $a_n$, $b_n$ as in the statement of the proposition, then  $\lim_{n\to\infty}\p(a_n(M_n-b_n)\leq y)=\e^{-\e^{-y}}$, which is the Gumbel distribution. 
\hfill$\square$

\begin{remark}
We observe that we can only do this computations because $\theta$ behaves exactly as the Lebesgue measure. Moreover and by (\ref{KK}) the stationary measure will be absolutely continuous; in this case the inverse technique  will not give any interesting information on the local dimension of the stationary measure, which will be integer;  instead we could use the existence of the limiting Gumbel law to detect the support of such a measure and how it depends on the intensity of the noise. Moreover we could compare the additive noise with the observational noise at different values of the noise to see how they affect the orbits of the original system:  we will report on these numerical computations in  the next sections.

\end{remark}

\section{Contractive maps: randomly applied stochastic perturbation}\label{CONTLASNOISE}
We study in this section and in the next one extreme value theory for contractive maps perturbed in two different ways. For the first class of perturbed maps, we will state rigorous results concerning the computations of the scaling coefficients $a_n$, $b_n$ and of conditions $D_2$ and $D'$. Moreover we will exhibit a point with an extremal index less than one. \\ We will succesively perturb contractive maps with additional noise giving in this case numerical evidence of existence of extreme value laws and of extremal indices: this will be done in Section \ref{SN}.\\

We consider here the random perturbation introduced in (\ref{RASP}), which were used in \cite{LM} as an illustration of the theory of constrictive transfer operators. This theory is an alternative to the standard treatment of the Perron-Frobenius operator based on quasi-compactness. In our present case the Perron-Frobenius operator admits a very explicit representation and this will allow us to to find an  expression for the levels sequence $(u_n)_{n\in\N}$ as well as verify the conditions $D_2(u_n)$ and $D'(u_n)$.\\

Let us therefore consider the map  $f$ defined on the unit interval $I=[0,1]$ by\footnote{We use here this map instead of $S(x)=\alpha x+\beta$ to simplify the computations; the theory and the results remain the same provided $\alpha+\beta\le 1$ and the map is therefore well defined over all the interval. Otherwise, namely if $\alpha+\beta>1$, we should consider a discontinuous map over the interval, and the techniques developed  in this paper should be modified. We believe this will lead to more elaborated statistical properties and we deserve to investigate them in the future.}
\[
    S(x)=\alpha x,\quad \alpha\in(0,1).
\]
We perturb it according to eq. (\ref{RASP}) giving rise to the family of random maps defined for each $n\ge 1$ by
\[
f_{\omega_n}(x)= \eta_nS(x)+(1-\eta_n)\xi_n,\quad\forall\,x\in I,
\]
where $\omega_n=(\eta_n,\xi_n)$.\\

In order to obtain the stationary measure $\mu_\eps$, let us introduce the random Koopman operator $U_{\eps}:L^{\infty}\to L^{\infty}$ defined for all $\phi\in L^{\infty}$\footnote{From now on $L^1$ and $L^{\infty}$ will be referred to the Lebesgue measure $m$ and the integral with respect to the latter will be denote as $\int (\cdot)\ dx.$} by
\[
U_{\eps}\phi(x):= \int \phi(f_{\omega}(x))d\theta_\varepsilon.
\]

The random transfer operator  $P_\eps$ is the adjoint operator of $U_\eps$; if we denote $P$  the transfer operator associated to $f$ and $\overline{\psi}= \int \psi(y)dy$, with $\psi\in L^1,$ then we have
\begin{equation}\label{PFCONT}
P_{\eps}\psi(x)= (1-\eps)P\psi(x)+\eps \overline{\psi}.
\end{equation}
The stationary measure $\mu_\eps$ verifies $\int \phi(x) d\mu_\eps=\int U_{\eps}\phi(x) d\mu_\eps$ and in our case is  given by $\mu_\eps=h_\eps m$ where $h_\eps\in L^1$ is a density such that $h_\eps=P_{\eps}h_\eps$. Such a density exists and is given by \cite{LM}:
\begin{equation}\label{DENS}
h_\eps=\eps \sum_{k=0}^{\infty}(1-\eps)^kP^k{\bf 1}.
\end{equation}\\
Our next step will be to get directly the linear scaling parameters $a_n$ and $b_n$, allowing to prove the convergence towards the Gumbel's law of the maxima of the process $X_n(x,\o) :=-\log(|f_{\o}^n(x)-z|)$. We first observe that 
the density $h_\eps$ could be explicitely computed and reads  for the the map $f(x)=\alpha x$ with $0<\alpha<1$ \cite{FFGV}:
\[
h_\eps(x)= \eps\sum_{k=0}^{p-1}\frac{(1-\eps)^k}{\alpha^k}\quad \forall\, x\in (\alpha^p, \alpha^{p-1}], \ p\ge 1.
\]
Notice that the density is bounded for $(1-\eps)< \alpha$.

To find a sequence $(u_n)_{n\in\N}$ satisfying \eqref{eq:un} we need to compute
$n\mathbb{P}(X_0>u_n).$ 

\begin{proposition}{\bf \cite{FFGV}}
Let $\tau>0$, $y=-\ln(\tau)$ and $u_n=\frac{y}{a_n}+b_n$ with
\[
a_n=1\qand \ b_n= \log\left(2n\eps\sum_{k=0}^{p-1}\frac{(1-\eps)^k}{\alpha^k}\right)\qquad\forall\,n\in\N,
\]
If:\\
(i) $z\neq 0$ on the interval but not in the countably many discontinuity points of $h_\eps$, namely $z\notin \cup_{j\in \mathbb{N}}\{\alpha^j\}$;\\
(ii) $p\geq 1$ such that $z\in (\alpha^p, \alpha^{p-1})$ and $n$  is large enough such that the ball $B(z,e^{-u_n})\subset(\alpha^p, \alpha^{p-1})$, then:
$$
n\mathbb{P}(X_0>u_n) =\tau
$$

\end{proposition}

The next step will be  to  check the conditions $D_2(u_n)$ and $D'(u_n)$. To verify condition $D_2(u_n)$ we need to show that for specifics observables $\phi\in L^{\infty}$ and $\psi\in L^1$  the correlation
 $$
 Cor_m(\phi,\psi,n) :=\left|\int  U^n_\eps(\phi(x))\psi(x)d\mu_\eps-\int \phi(x)d\mu_\eps \int \psi(x)d\mu_\eps\right|
 $$
\begin{equation}\label{COR}
\left|\int\int  \phi(f^n_{\o}(x))\psi(x)d\mu_\eps d\theta_\eps^\N-\int \phi(x)d\mu_\eps \int \psi(x)d\mu_\eps\right|
\end{equation}
decay sufficiently fast with $n$. 
At this regard we have:
\begin{proposition}{\bf \cite{FFGV}}
If $\phi\in L^{\infty}$ and $\psi\in L^1\cap L^\infty$, then 
\[
Cor_m(\phi,\psi ,n)\leq 2(1-\eps)^n||\phi||_{L^\infty} ||\psi h_\eps||_{L^1}.
\]
\end{proposition}
We notice that condition $D_2$ requires that $\psi$ and $\phi$ are characteristic functions of measurable sets. Condition $D'$ needs to control short returns in the ball around $z$. We can prove that it holds for $z\neq 0$ and the proof is basically based on the fact that the image of the ball  does not intersect the ball itself for large $n$. Instead, whenever $z=0$ an extremal index appears on the limiting law for the distribution of the maxima. We summarised it into the next proposition
\begin{proposition}{\bf \cite{FFGV}}
For the map $S(x)=\alpha x, \ \alpha\in (0,1)$ perturbed with the noise (\ref{RASP}), and by considering the observable $X_0(x)=-\log(|x-z|),$ conditions $D_2$ and $D'$ hold.\\ If $z=0$, conditions $D_p(u_n)$, $D'_p(u_n)$ (introduced in \cite{FFT2}) hold which implies the existence of an extremal index less than 1, that is given by $\eps$. 
\end{proposition}

\section{Numerical computations}\label{SN}
This chapter contains a comparison between the numerical effects of random transformations and observational noise on two famous attractors.  The starting point of our analysis will be the fact that by perturbing the map with additive and uniform noise,  then the linear scaling parameter  $b_n$ of the Extreme Value Theory is expected to behave as

\begin{equation}
b_n \sim \frac1d \ \log(n \epsilon^{D-d}).
\label{superformula}
\end{equation}
where $d$ is the ambient space dimension and $D$ the the Hausdorff dimension of the stationary measure. Since the stationary measure will be smooth (see (\ref{KK})), the quantity $b_n$ will only depend on the phase space dimension. The situation is completely different if we perturb with observational noise, because in this case the above formula remains true with $D$ as an estimator of the Haudsdorff dimension of the invariant (SRB) measure.

\subsection{Lozi map}\label{SECLOZI}

\begin{figure}[ht]
\centering
\includegraphics[width=100mm]{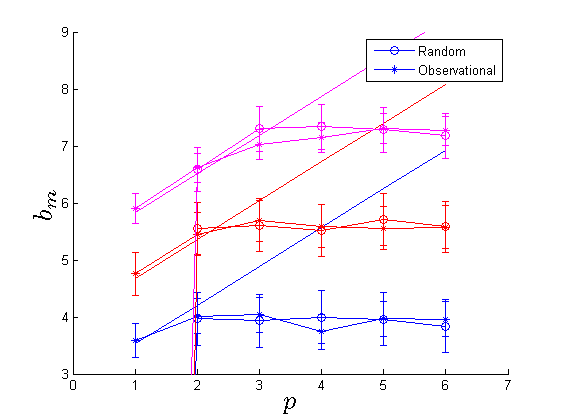}
\caption{Normalizing sequence $b_n$ vs intensity of the noise in terms of $p$  (we recall that $\epsilon=10^{-p}$) for the Lozi map  (Eq. \ref{lozi}).  Errorbars display the average of $b_n$ over 30 realizations and the standard deviation of the sample. Solid lines the theoretical values. the blue, red and magenta curves respectively refers to $n=1000,10000,30000$. The points $z$ are randomly chosen on the attractor. Observational noise is represented by stars, random transformation by circles.}
\label{lozifig}
\end{figure}

Let's consider  the Lozi map:
\begin{equation}
\begin{array}{lcl}
x_{t+1}&=&y_t +1 -a |x_t|\\
y_{t+1}&=&b x_t\\
\end{array}
\label{lozi}
\end{equation}
for which we consider the classical set of parameter   $a=1.7$ and $b=0.5$. Young   proved the existence of the SRB measure for the Lozi map and found the value $D=1.40419$ for the Hausdorff  dimension of the measure     by computing the Lyapunov exponents  and using a Kaplan-Yorke like formula.   The experiments we performed  consist of computing $30$ realizations of the maps perturbed with  observational and additive noise. Again, we  fit the maxima of the observable $$w=-\log(dist(\vec{x}_t, \vec{\zeta}))$$ to the Gumbel distribution  by using the $L$-moments   procedure and compare the values of $b_n$ obtained experimentally to the theoretical ones stated in  Eq. \ref{superformula}  (see \cite{FFTV} for details on the inference procedure).  We report   in Fig.~\ref{lozifig}   the results for three different bin lengths $n=1000,10000,30000$. For all the computations we fix the number of maxima to perform the fit equal to  $1000$ which has been proven to be a reasonable value in \cite{FLTV}. Agreement with the theoretical  $b_n$ represented by the solid straight lines, is found for small   values of $p$ which means for large $\epsilon$, since $\epsilon=10^{-p}$ . This is true only for the observational noise. In fact, for the random transformation, at $p=1$, points escape from the attractor and the statistics diverge. For lower noise intensities $p>1$, the $b_n$ have the same behavior for both random transformations and the observational noise: first they follow the theoretical prediction (solid line) for perturbed systems. Then, when the noise intensity is low, they approach a plateau which corresponds to the deterministic limit. This plateaus is achieved at higher noise intensities for shorter time series ($n=1000$, the blu lines), whereas if we consider the dynamics for longer times ($n=30000$) we can follow the prediction to lower intensities of the noise $p\simeq3$.\\

\begin{figure}
\centering
\includegraphics[width=150mm]{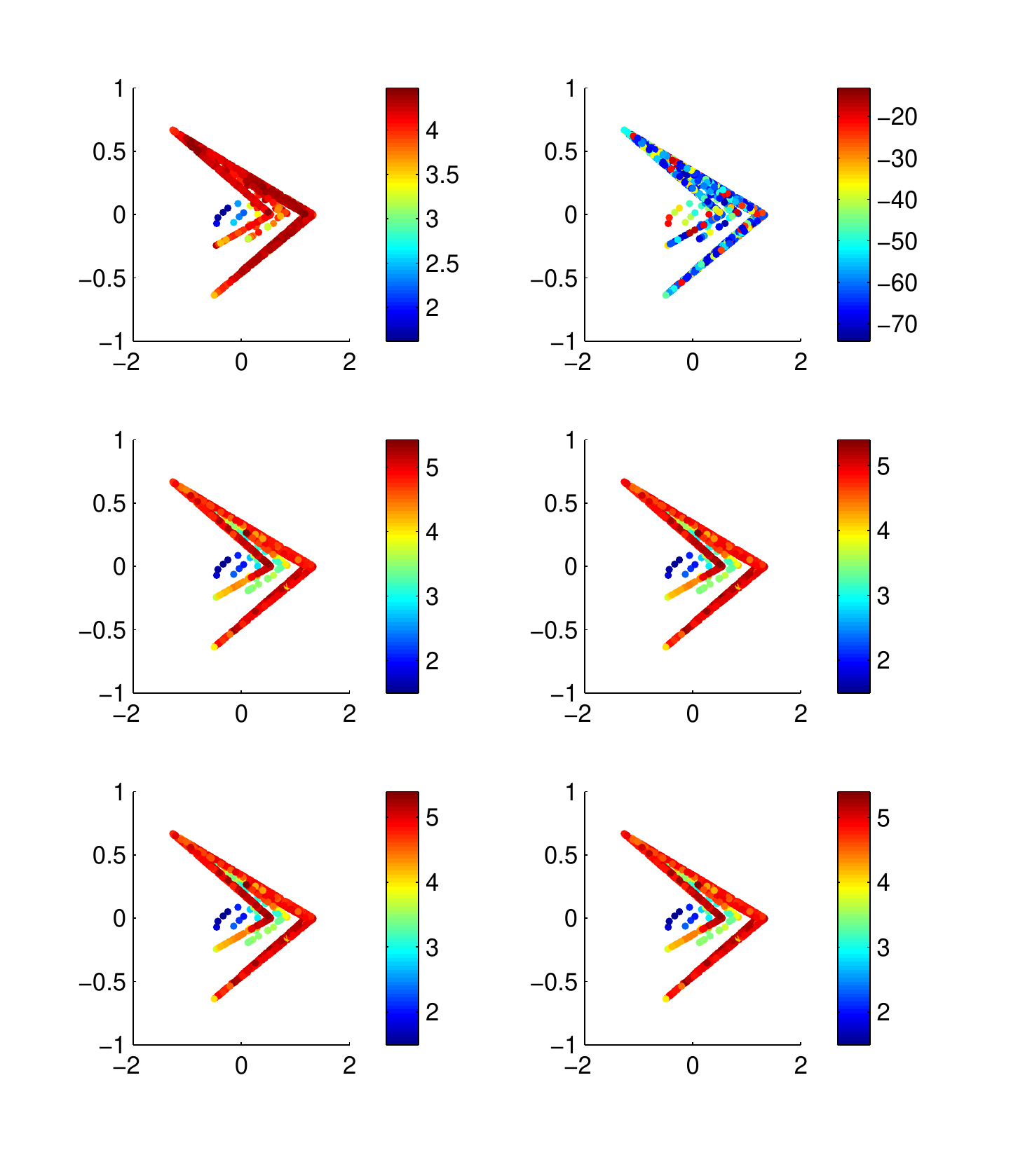}
\caption{Map of $b_n$ (colorscale)  for the Lozi system  (Eq. \ref{lozi}). The left plots refer to the observational noise, the right plots  to random transformations. Top panels: $p$=1, Central panels $p=3$, Lower panels $p=5$. }
\label{bm}
\end{figure}

\begin{figure}
\centering
\includegraphics[width=150mm]{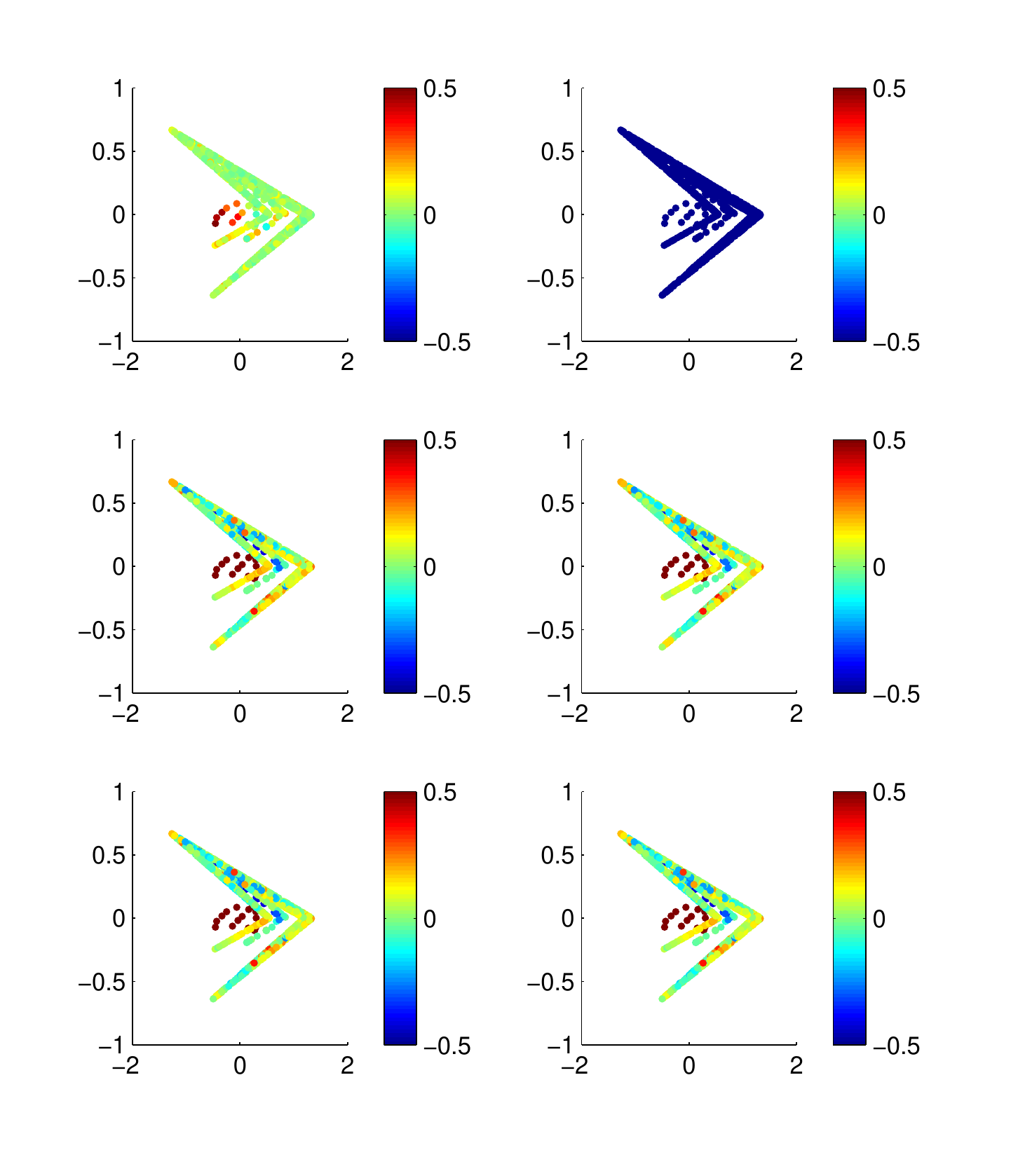}
\caption{Map of distances from the Gumbel law (colorscale)  for the Lozi system  (Eq. \ref{lozi}). The left plots refer to the observational noise, the right plots  to random transformations. Top panels: $p$=1, Central panels $p=3$, Lower panels $p=5$. }
\label{csi}
\end{figure}

Let's now produce a map of the $b_n$ (Fig. \ref{bm}) and of the distance from the Gumbel law \footnote{computed as deviations of the shape parameter of the EVLs fit from the expected one} (Fig. \ref{csi}) when the noise intensity $p$ is changed. In this case we will study local results, which depend on the point $\zeta$ chosen and on the intensity of the noise.
Let's begin with the analysis of $b_n$ ( Fig. \ref{bm}). The value $n=1000$ is fixed for all the figures, what we change is  the type of perturbation (left panels for the observational noise and right panels for random transformations) and the intensity of the noise (Upper panels: $p$=1; Central panels $p=3$; Lower panels $p=5$). This means that  the intensity of the noise decreases from the top to the bottom of the figures given that $\epsilon=10^{-p}$. A first remarkable difference is the divergence of the values of $b_n$ for the random transformations at $p=1$. We have already seen this in the previous analysis but here we can remark that this effect is $\zeta$-dependent as we obtain a spectrum of different $b_n$. The values of $b_n$ are, instead, rather uniform for decreasing noise intensities $p=3$ and $p=5$ and show a substantial identity between the observational noise and the random transformations. \\

Similar conclusions hold for the analysis of distances from the Gumbel law reported in Fig. \ref{csi}. Here for $p=1$ (top panels) the observational noise (left) show that all the points have more or less well converged to the Gumbel law (0 values) whereas great divergences appear for the random transformations (right). The differences become less and less evident as far as the noise intensity is decreased ($p$=3 central panels and $p=5$ lower panels). In the weak noise limit, one can clearly see that there are points which converge better to the Gumbel law and points that have a worst convergence (remind that $m$ is fixed to be 1000 independently on the point chosen). Remarkably, not many differences appear between observational noise and random transformations when the noise is small ($p$=3) and in the deterministic limit $p=5$.\\

\subsection {H\'enon map}\label{SECHENON}

\begin{figure}
\centering
\includegraphics[width=150mm]{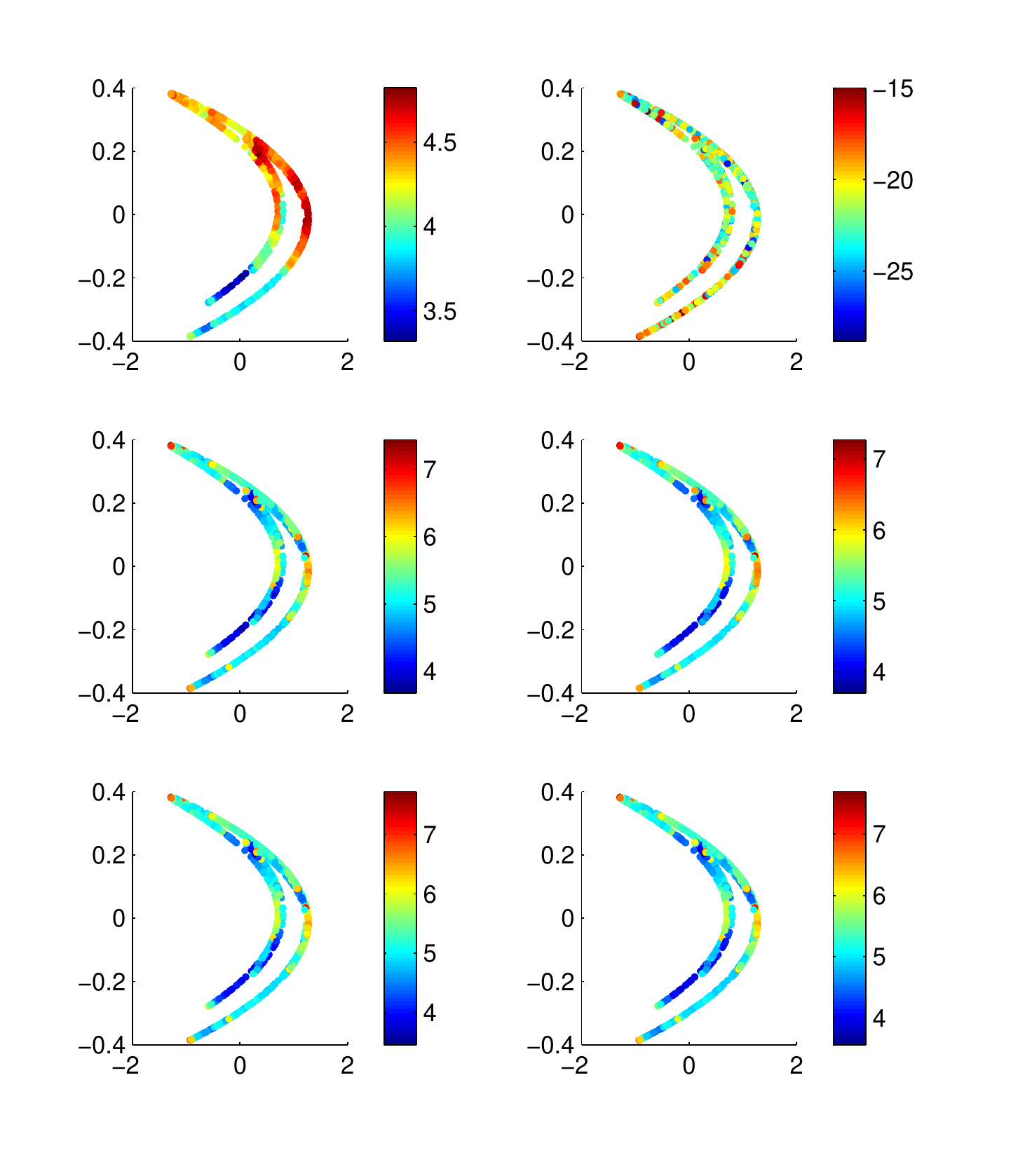}
\caption{Map of $b_n$ (colorscale)  for the Lozi system  (Eq. \ref{henon}). The left plots refer to the observational noise, the right plots  to random transformations. Top panels: $p$=1, Central panels $p=3$, Lower panels $p=5$. }
\label{bmh}
\end{figure}

\begin{figure}
\centering
\includegraphics[width=150mm]{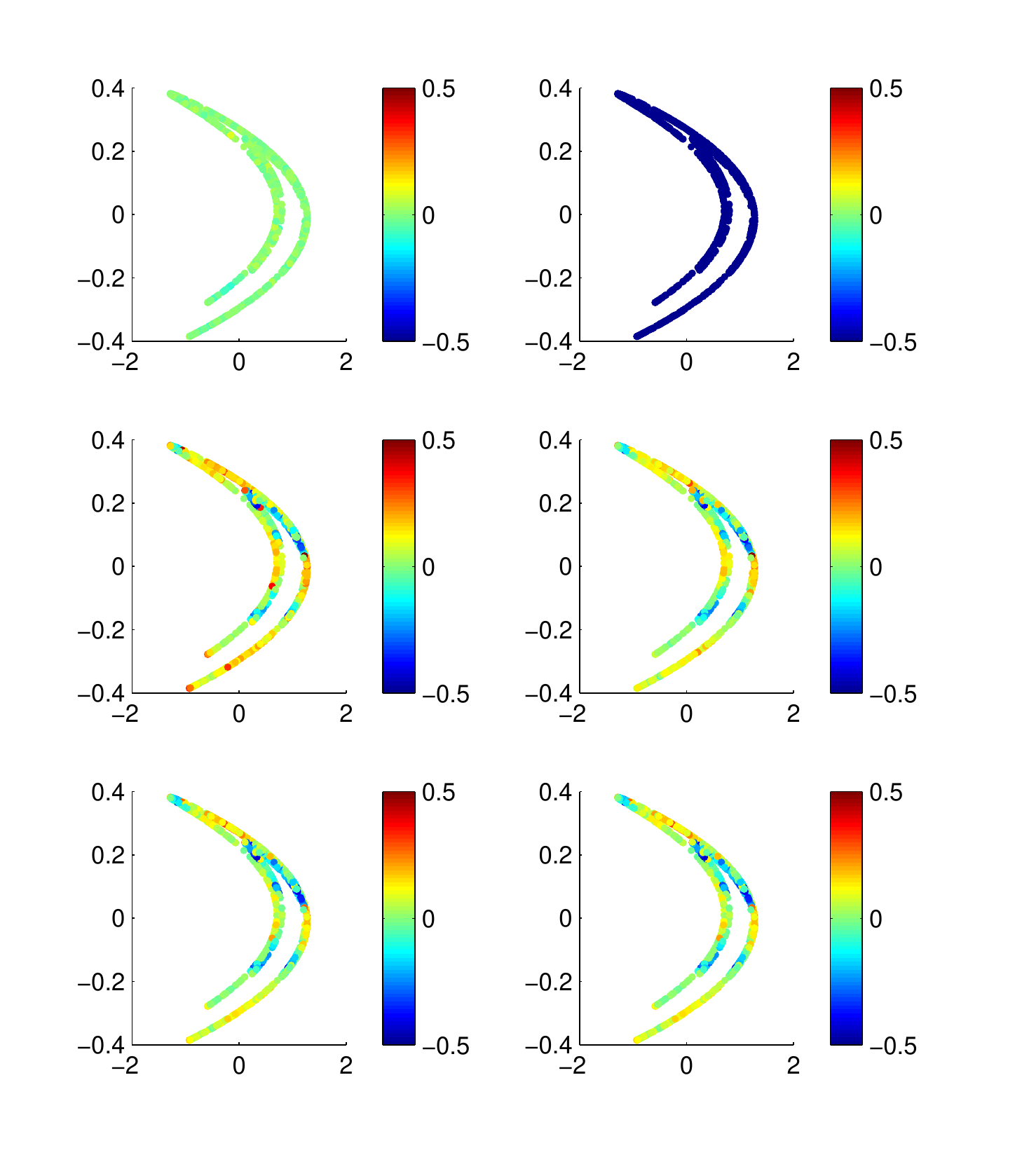}
\caption{Map of distances from the Gumbel law (colorscale)  for the Lozi system  (Eq. \ref{henon}). The left plots refer to the observational noise, the right plots  to random transformations. Top panels: $p$=1, Central panels $p=3$, Lower panels $p=5$. }
\label{csih}
\end{figure}

Similar results can be found also for the H\'enon map:

\begin{equation}
\begin{array}{lcl}
x_{t+1}&=&y_t +1 -a x^2_t\\
y_{t+1}&=&b x_t\\
\end{array}
\label{henon}
\end{equation}

for which we have taken the classical parameters $a=1.4$ and $b=0.3$. Analytical results are harder to get and estimations of the dimensions are known only numerically \cite{grassberger1983}. We perform the same analysis as for the lozi maps, reporting the results in Figs. \ref{bmh}-\ref{csih}. Results follow the ones found for the Lozi map: for the observational noise, fits converge better to the Gumbel law  whereas diverge for the additive noise (respectevely Top-Left and Top-Right panels of Fig. \ref{csih}). The same happens for the $b_n$. As it is happen for the Lozi dynamics, large additive noise let the system escape from the H\'enon attractor causing divergent extreme value laws. For smaller values of the noise (central and bottom panels of Figs. \ref{bmh} and \ref{csih}), distances from the Gumbel law increase depending on the point $\zeta$ chosen and values of $b_n$ attain the deterministic limit.\\

We can summarise the analysis for the Lozi and H\'enon map as follows: in the limit of small noise, no substantial differences appear between the random transformations and the observational noise. However, for large noise (or even for average noise but if we wait enough time) we observe jumps outside the  attractors for the random transformations which we do not observe for the observational noise, as we are forced to stay close to the attractor. Remarkably, the observational noise is more \textit{stable} than the random transformation and by using it one can obtain an excellent convergence to the Gumbel law at large noise intensity without risking to escape from the attractor, as instead it happens for the random transformations. \\

We can therefore suggest a general strategy to study strange attractors by using the observational noise. In fact,  for large noise intensities we can obtain better fit to the Gumbel law and retain the information on the local properties of the measure in the expression of $b_n$. The distance from the Gumbel law will be an indication on how good is our fit and therefore on how good are the estimates of the value of the local dimension inferred by $b_n$.

\subsection{Piecewise contracting maps with additive noise}\label{PCM}
Until now, the existence of an EVL has been principally established for expansive maps and rotations of the circle perturbed with additive noise \cite{AFV, FFTV}. We have shown in Section \ref{CONTLASNOISE} that pure contractions can also admit an EVL, provided they are suitably perturbed. Here we discuss the case of additive noise applied to piecewise contracting maps already studied, e.g. in \cite{C, K80}.

The simplest piecewise contracting map is defined on the unit interval $I$ (or the circle, also denoted $I$), by
\begin{equation}\label{PCMI}
f(x)=\alpha x +\beta \mod 1\quad\forall\,x\in I,
\end{equation}
where $\alpha$ and $\beta$ belong to $(0,1)$. If $\alpha+\beta>1$, then $S$ is discontinuous at the point $\gamma=(1-\beta)/\alpha$ (when considered as a map of the interval) and each restriction of $S$ to one of the open interval $[0,\gamma)$ and $[\gamma, 1)$ is a contraction. It has been shown that for generic values of the parameters $\alpha$ and $\beta$, any orbit of $S$ is attracted by a periodic orbit, whose period can be arbitrarily big (depending on the parameters). On the other hand, for an uncountable number of pairs $(\alpha,\beta)$ the attractor is a Cantor set supporting a minimal dynamics.

When perturbing with additive noise, we should consider random orbits $(x_n)_{n\in\N}$ satisfying
\begin{equation}\label{E}
x_{n+1}=f(x_n) + \omega_n \mod 1\quad\forall\,n\in\N,
\end{equation}
where $f(x)=\alpha x+\beta$ for all $x\in I$, and the quantities $(\omega_n)_{n\ge1}$ are i.i.d. random variables with values in a small interval, say in $\Omega_{\eps}:=[-\eps, \eps], 0<\eps<1,$ with a common distribution $\theta_{\eps}$ given by a density $g$, namely $d\theta_{\eps}(\omega)=g(\omega)d\omega$, with $\int_{-\eps}^{\eps}g(\omega)d\omega=1.$ We notice that whenever $I$ is the (unit) interval, even if $\alpha+\beta<1$, the strength  of the noise, namely $\eps$, should be small enough in such a way the image of $I$ is still in $I$. This problem disappears provided we take the mod-$1$ operation.

It is easy to check that the random Perron-Frobenius operator $P_{\eps}$ introduced in the previous section, now acts on the $L^1$ real valued functions (w.r.t. the Lebesgue measure $m$) $\psi$ on $I$ as
\begin{equation}\label{P}
P_{\eps}\psi(x)= \int_I \psi(y) g(x-f(y)) dm(y)
\end{equation}
If we  now add the additional  assumption  that the first moment of $g$ is finite, then  the operator $P_{\eps}$ becomes  {\em weakly constrictive} according to the definition proposed in \cite{LM}. We defer to \cite{LM} for the precise definition; what is important to retain here is that there exists $r$ measurable functions $g_i$  with disjoint supports which are cyclically permutated by ${\cal P}$, namely  $P_{\eps}g_i=g_{w(i)},$  where $\{w(1),\dots, w(r)\}$ is a permutation of $\{1,\dots,r\}$. Moreover if $\psi\in L^1$ then $P_{\eps}^n \psi$ converges in the $L^1$ norm to $\sum_{i=1}^r g_{w^n(i)} \int \psi(x)k_i(x)dx,$ where $\{w^n(1),\dots,w^n(r)\}$ is a permutation of $\{1,\dots,r\}$, and  the $k_i$ are suitable $L^{\infty}$ functions.
The last two items justify the appellation of {\em asymptotically aperiodic} given to the sequence $(P_{\eps}^n \psi)_{n\in\N}.$

\begin{figure}
\centering
\includegraphics[width=.5\textwidth]{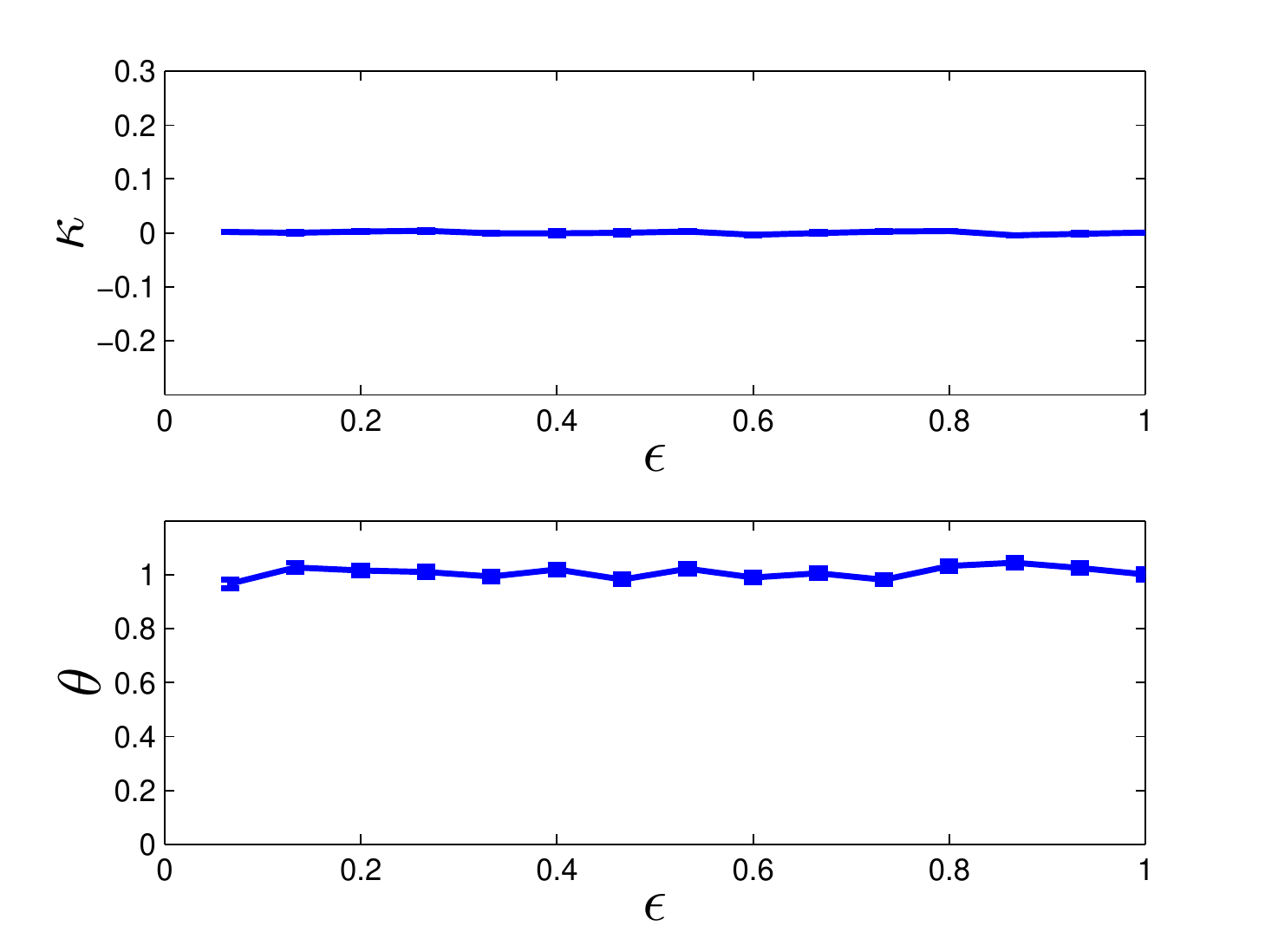}\includegraphics[width=.5\textwidth]{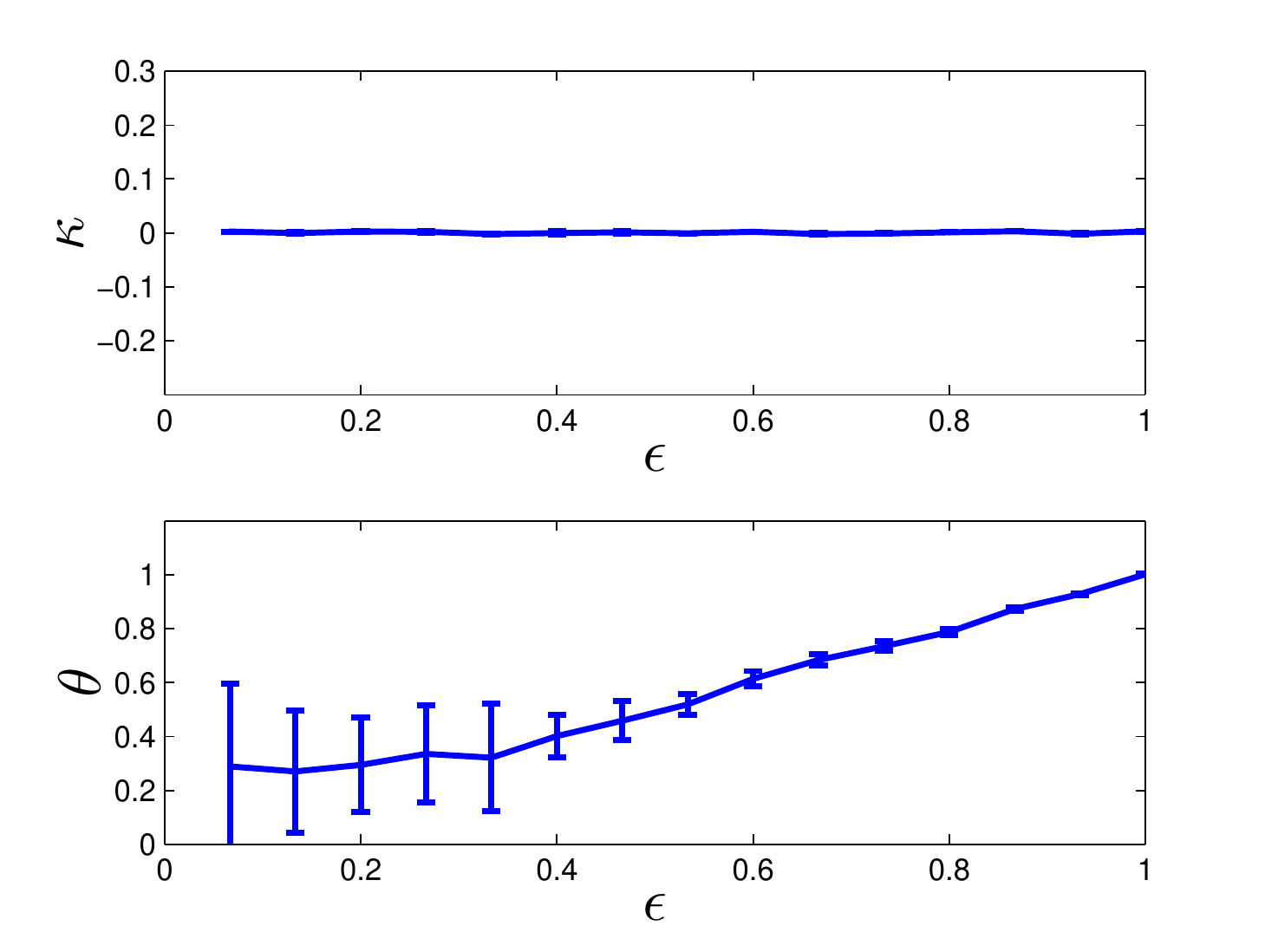}
\caption{Upper panels: shape parameter $\kappa$ VS noise intensity $\epsilon$ for the map \eqref{E}. Lower panels: Extremal index $\vartheta$ VS noise intensity $\epsilon$ for the map \eqref{E}. Left panels: $\alpha=1/3$, $\beta=7/8$,  $n=1000$.  Right panels:  $\alpha= 1/2$,  $\beta= 17/30$ $n=1000$. The errorbars represent the standard deviation of the mean.}
\label{contra}
\end{figure}

It was pointed out by Lasota and Mackay that a small stochastic perturbation of the transformation (\ref{PCMI}) with an aperiodic unperturbed asymptotic dynamics (i.e, supported by a Cantor set), is enough for the random dynamical system \eqref{E} to have a transfer operator \eqref{P} asymptotically periodic.
Contrarily to the noise investigated in Section 4, we are not able to produce rigorous results  to show in particular that conditions $D_2$ and $D'$ are satisfied, but some information is still available. First of all we observe that the convex combination of the $g_i$ given by the spectral decomposition and with equal weights $\frac1r$ gives a stationary measure. In particular it has been proved by Lasota and Mackay that such a measure is mixing if and only if $r=1.$  This seems the case for the map $S$ with $\alpha=1/3$, $\beta=7/8$ and the noise $\eps$ large enough. The random orbits of arbitrarly chosen initial conditions seem to distribute on the whole circle suggesting the presence of only one fixed point for $P_\eps$ and therefore mixing. We do not have any means to compute the rate of decay, but the GEV statistics effectuated for the usual observable shows a pretty good convergence towards the Gumbel's law. This is what is represented in  Fig.~\ref{contra} (upper-left panel) where the shape parameter $\kappa$ is plotted against the noise intensity $\epsilon$. For the range of values investigated  there are no substantial deviation from 0, that is the Gumbel law. Error-bars have been computed as the standard deviation of an ensamble of 30 realizations of the map, with $n=1000$.

As mention earlier, another interesting case is given by a particular choice of the parameters namely $\alpha= 1/2,  \beta= 17/30$ and the random variables $\omega$ distributed uniformly in the interval $[0, \eps]$ with $\eps=1/15.$ This example has been also investigated by Lasota and Mackay; these values of parameters are close to (irrational) values for which   in the absence of noise the orbits $(x_n)_{n\in\N}$ are not periodic and the invariant limiting set $\cap_{k=0}^{\infty}S^kI$ is a Cantor set.
The presence of noise will induce asymptotic periodicity in the sense that we detect a sequence of densities which is asymptotically periodic with period  $r=3$.

This is particularly evident if we perform an histogram of the frequency of visits of random orbits on the circle. We have therefore taken the center $z$ of our target balls in the supports of these densities and checked if there is presence of extreme value statistics. Numerical computations suggest the presence of an extremal index equal to $1/3$ which could be related to the order of asymptotic periodicity. These results are shown in  the right panels of Fig.~\ref{contra}. The upper plot is, as before, the shape parameter $\kappa$ against the noise intensity $\epsilon$. The smallest $\epsilon$ plotted is 1/15 so that we can remark how the convergence to the Gumbel law is good and does not change with increasing noise. The extremal index  (bottom-left panel of the same figure) shows instead a clear dependence on $\epsilon$. For $\epsilon \to 1/15$ a good convergence towards the prediction $\vartheta=1/3$ is achieved, whereas for $\epsilon \to 1$, also $\vartheta \to 1$ as we expect from the theoretical arguments of \cite{FV1}.

\medskip\noindent\textbf{Acknowledgement}.
SV was supported by the ANR- Project {\em Perturbations} and by the project ``Atracci\'on de Capital Humano Avanzado del Extranjero" MEC 80130047, CONICYT, at the CIMFAV, University of Valparaiso. JMF was partially supported by FCT (Portugal) grant SFRH/BPD/66040/2009, by FCT project PTDC/MAT/120346/2010, which is financed by national and by the European Regional Development Fund through the programs  FEDER and COMPETE . JMF was also supported by CMUP, which is financed by COMPETE and by the Portuguese Government through FCT, under the project PEst-C/MAT/UI0144/2013. S Vaienti thanks R Aimino for useful discussions. DF was supported by a CNRS postdoctoral grant.
\bibliographystyle{unsrt}


\end{document}